\documentstyle[aps,preprint]{revtex}

\newcommand{\beq}{\begin{equation}}
\newcommand{\eeq}{\end{equation}}
\newcommand{\beqa}{\begin{eqnarray}}
\newcommand{\eeqa}{\end{eqnarray}}

\begin{document}

\tighten

\def\epsilondelta{\forall \epsilon > 0 \exists \delta > 0}
\def\rl{\rangle \langle}
\def\openone{\leavevmode\hbox{\small1\kern-3.8pt\normalsize1}}
\def\RR{{\rm I\kern-.2emR}}
\def\tr{{\rm tr}}
\def\ce{{\cal E}}
\def\cc{{\cal C}}
\def\ci{{\cal I}}
\def\cd{{\cal D}}
\def\cb{{\cal B}}
\def\cn{{\cal N}}
\def\ct{{\cal T}}
\def\cf{{\cal F}}
\def\ca{{\cal A}}
\def\cg{{\cal G}}
\def\ch{{\cal H}}
\def\cv{{\cal V}}
\def\cc{{\cal C}}
\def\rhon{\rho^{\otimes n}}
\def\on{^{\otimes n}}
\def\pn{^{(n)}}
\def\pnp{^{(n)'}}
\def\tcd{\tilde{\cal D}}
\def\tcn{\tilde{\cal N}}
\def\tct{\tilde{\cal T}}
\def\id{\frac{I}{d}}
\def\pthang{\frac{P}{\sqrt{d \tr P^2}}}
\def\psirq{\psi^{RQ}}
\def\rhorq{\rho^{RQ}}
\def\rhorqp{\rho^{RQ'}}
\def\ra{\rangle}
\def\la{\langle}
%%%IEEE proof style
\def\QED{\mbox{\rule[0pt]{1.5ex}{1.5ex}}}
\def\proof{\noindent\hspace{2em}{\it Proof: }}
\def\endproof{\hspace*{\fill}~\QED\par\endtrivlist\unskip}

%%%new commands from john & barbara
\newcommand{\half}{\mbox{$\textstyle \frac{1}{2}$} }
\newcommand{\ket}[1]{| #1 \rangle}
\newcommand{\bra}[1]{\langle #1 |}
\newcommand{\proj}[1]{\ket{#1}\! \bra{#1}}
\newcommand{\outerp}[2]{\ket{#1}\! \bra{#2}}
\newcommand{\inner}[2]{ \langle #1 | #2 \rangle}
\newcommand{\melement}[2]{ \langle #1 | #2 | #1 \rangle}

\date{1999 July 7}
\title{Quantum Probability from Decision Theory?}

\author{H.~Barnum,$^{(1)}$ C.~M.~Caves,$^{(2)}$ J.~Finkelstein,$^{(3)}$ 
C.~A.~Fuchs,$^{(4)}$ and R.~Schack$^{(5)}$}

\address{
$^{(1)}$School of Natural Science and Institute for Science and
Interdisciplinary Studies (ISIS),\\ Hampshire College, Amherst, MA
01002, USA \\ 
$^{(2)}$Center for Advanced Studies, Department of Physics
and Astronomy,\\ University of New Mexico, Albuquerque, New Mexico
87131-1156, USA \\ 
$^{(3)}$Department of Physics, San Jos{\'e} State University,
San Jos{\'e}, CA 95192, USA \\
$^{(4)}$Norman Bridge Laboratory of Physics, 12--33,\\
California Institute of Technology, Pasadena, CA 91125, USA \\ 
$^{(5)}$Department of Mathematics,\\ Royal Holloway, University of
London, Egham, Surrey TW20 0EX, UK } \maketitle

\begin{abstract}\relax
In a recent paper, Deutsch \cite{Deutsch99} claims to derive the
``probabilistic predictions of quantum theory'' from the 
``non-probabilistic axioms of quantum theory'' and the 
``non-probabilistic part of classical decision theory.''  We show 
that his derivation fails because it includes hidden probabilistic 
assumptions.  
%that are masked by ambiguities in notation.
\end{abstract}

\section{Introduction}

In a recent paper, Deutsch \cite{Deutsch99} attempts to derive the
``probabilistic predictions of quantum theory'' from the 
``non-probabilistic part of quantum theory'' and what he views as 
the ``non-probabilistic part of classical decision theory.''  For
Deutsch this means the following.  The nonprobabilistic part of 
quantum theory is contained in the axioms that associate quantum states 
with rays in Hilbert space and observables with Hermitian operators;  
in particular, the eigenvalues of a Hermitian operator are the only 
possible results of a measurement of the corresponding observable, 
and if the quantum state is an eigenstate of the Hermitian operator, 
the eigenvalue is the certain result of a measurement of that observable.  
The relevant nonprobabilistic part of classical decision theory includes
the assumption that a rational decision maker orders all his preferences 
transitively---that is, if he prefers $A$ to $B$ and $B$ to $C$, he 
must also prefer $A$ to $C$.  From these assumptions, Deutsch seeks to 
derive, first, that quantum mechanics has a probabilistic interpretation 
and, second, that the quantum probability rule has the standard form 
of a squared inner product.  Deutsch describes his result as follows:
\begin{quote}\small\baselineskip=12pt
Thus we see that quantum theory permits what philosophy would hitherto 
have regarded as a formal impossibility, akin to `deriving an ought 
{}from an is', namely deriving a probability statement from a factual 
statement.  This could be called deriving a `tends to' from a `does'.
\end{quote}

We argue in this paper that Deutsch's derivation fails to achieve both
its goals.  First, as we discuss in Sec.~\ref{sec:assumptions}, 
the standard nonprobabilistic axioms of classical decision theory, which
include the assumption of (complete) transitive preferences, 
already ensure that 
the preferences can be ordered in terms of probabilities and utility 
functions \cite{Savage72,vonNeumann47}.  Second, as we detail in 
Sec.~\ref{sec:flaw}, Deutsch's derivation of the form of the quantum 
probability law is flawed because an ambiguity in his notation masks a 
hidden probabilistic assumption that is essential for the derivation.  

Despite the failure of Deutsch's derivation, we are sympathetic to 
the view that the meaning of probability in quantum mechanics is 
specified by its role in rational decision making.  Indeed, we believe 
that this view can help illuminate the very nature of quantum theory 
\cite{Caves00}.  We believe, however, that the primary technical 
machinery underlying this view is already provided by Gleason's 
theorem \cite{Gleason57}, an oft-neglected derivation of the quantum 
probability law.  We review the theorem in Sec.~\ref{sec:conc}.  Gleason 
assumes that observables are described by Hermitian operators, 
supplementing that only by the assumption that the results of 
measurements cannot always be predicted with certainty and that the 
uncertainty is described by probabilities that are consistent with 
the Hilbert-space structure of the observables.  From this he is able 
to derive both that the possible states are density operators and that 
the quantum probability law is the standard one.  Because Gleason's 
theorem gives both the state space of quantum mechanics and the 
probability rule, we believe it trumps all other derivations along 
these lines.

\section{Probabilities and decision theory}
\label{sec:assumptions}

Classical decision theory, formulated along the lines that Deutsch 
has in mind, envisions a rational decision maker, or agent, who is
confronted with a choice among various games \cite{Savage72,vonNeumann47}.  
Each game is described by a set of events labeled by an index $j$, which the 
agent believes will occur with probability $p_j$.  The value the agent 
attaches to an event within a given game is quantified by its utility 
$x_j$.  Decision theory seeks to capture the notion of rational decision 
making by positing that the agent decides among the games by choosing the 
one that has the largest expected utility,
\begin{equation}
\sum_jp_jx_j\;.
\end{equation}
A simple consequence of this framework is that an agent can give his
preferences among games a complete transitive ordering.

Deutsch extracts what he sees as the nonprobabilistic part of decision 
theory and applies it to quantum mechanics in the following way.  A 
game is again described by events, now interpreted as the outcomes 
of a measurement of a Hermitian operator that has eigenstates
$|\phi_j\rangle$.  The $j$th outcome has utility $x_j$.  
In place of the probabilities of classical decision theory, Deutsch
substitutes the normalized quantum state of the system in question,
\begin{equation}
|\psi\rangle=\sum_j\lambda_j|\phi_j\rangle\;.
\end{equation} 
Thus a quantum game, in Deutsch's formulation, is characterized by a 
quantum state and utilities that depend on the outcome of a measurement 
performed on that state.
%a utility operator.  
As the final part of his 
formulation, Deutsch defines the {\em value of a game}---the central 
notion in his argument---as ``the utility of a hypothetical payoff such 
that the player is indifferent between playing the game and receiving 
that payoff unconditionally.''  Deutsch does not assume that the value 
of a game is an expected utility, for that is precisely the probabilistic 
aspect of classical decision theory he wants to exclude from his 
formulation.  He does assume that the values are transitively ordered 
and that a rational decision maker decides among games by choosing the 
game with the largest value.  

Deutsch describes this in the following way:
\begin{quote}\small\baselineskip=12pt
On being offered the opportunity to play such a game at a given price, 
knowing $|\psi\rangle$, our player will respond somehow: he will either 
accept or refuse.  His acceptance or refusal will follow a strategy 
which, given that he is rational, must be expressible in terms of 
transitive preferences and therefore in terms of a value $\cv[\ket{\psi}]$ 
for each possible game.
\end{quote}
Notice that Deutsch denotes the value of a game without explicit reference 
to the utilities and the corresponding eigenstates, which partially define
the game.  We prefer a more explicit notation.  First we define a Hermitian
{\it utility operator}
\begin{equation}
\hat X=\sum x_j|\phi_j\rangle\langle\phi_j|
\;.
\end{equation} 
We now can denote the value of a game more explicitly as 
$\cv(|\psi\rangle;\hat X)$, which includes both defining features 
of a game, the quantum state $|\psi\rangle$ and the utility operator 
$\hat X$.  Our notation serves its purpose in Sec.~\ref{sec:flaw}, 
where it helps to ferret out a flaw in Deutsch's derivation.

Before turning to such details of Deutsch's argument, we consider a more 
fundamental issue.  Deutsch's attempt to derive the probabilistic 
interpretation of quantum mechanics from purely nonprobabilistic 
considerations must fail, because his assumption of complete transitive 
preferences is tantamount to assuming probabilities at the outset.  The
conventional understanding of preferences---making decisions in the face 
of uncertainty---already hints strongly that probabilities will be an 
essential tool in any decision theory.  Indeed, this is the import of a 
fundamental result of the theory \cite{Savage72,vonNeumann47}: if one assumes
complete transitive preferences among games 
along with standard nonprobabilistic axioms, one can determine 
simultaneously utility functions and sets of probabilities, such that 
the agent's behavior is described as maximizing expected utility. 
If the preferences among games are quantified by a value function $\cv$, 
then for each game there exist probabilities $p_j$ and 
transformed utilities $F(x_j)$, where $F$ is a strictly increasing function,
such that expected utility gives the same ordering:
\begin{equation}
\cv(|\psi\rangle;\hat X)=F^{-1}\Biggl(\sum_jp_jF(x_j)\Biggr)\;.
\end{equation}
The crucial point is that this intimate relationship between preferences
and probabilities is purely classical, having nothing to do with quantum
mechanics.

Although Deutsch's argument fails to exclude {\it a priori\/} probabilistic 
considerations, it might nevertheless provide a derivation of the specific 
form of the quantum probability law.  To assess this possibility, we turn 
now to Deutsch's specific argument.

\section{Examination of Deutsch's derivation} 
\label{sec:flaw}

In this section we examine the derivation of what Deutsch terms the  
``pivotal result'' of his argument, 
\begin{equation}
\cv\!\left({1\over\sqrt2}\big(|\phi_1\rangle+|\phi_2\rangle\big);\hat X\right)=
{1\over2}(x_1+x_2)\;;
\label{eq:pivotal}
\end{equation}
that is, that the value of a game in which the quantum state is an 
equal linear combination of two eigenstates of the utility operator is
the mean of the utilities.  The derivation of the pivotal result is
contained in Deutsch's Eqs.~(D7)--(D11) and related textual material.   
(Here and throughout we refer to equations in Deutsch's paper by prefixing
a D to the equation number.)  We comment briefly on later steps in 
Deutsch's argument at the end of this section.

Deutsch uses a notation for the value of a game that makes no explicit
reference to the utility operator.  Furthermore, he employs a notational
convention, often used in physics, whereby an eigenvector of an 
operator---in this case the utility operator---is labeled by its 
eigenvalue---in this case the utility itself.  This labeling can cause 
confusion when games involving different utility operators are under 
consideration, as in the argument examined in this section.  The
resulting ambiguity leads Deutsch to accidentally equate the value of 
two games whose value cannot be shown to be equal without some additional 
assumption.  Identifying this hidden assumption is the goal of this 
section.

In deriving the pivotal result, Deutsch posits two properties of the 
quantum value function.  The first, given in Eq.~(D8), we call the 
{\it displacement property}.  Written in our notation, this property 
becomes 
\begin{equation}
{\cal V}\!\left( \sum_j\lambda_j|\phi_j\rangle;\sum_j
(x_j+k)|\phi_j\rangle\langle\phi_j|\right)
 = k + {\cal V}\!\left(
\sum_j\lambda_j|\phi_j\rangle;\sum_jx_j|\phi_j\rangle\langle\phi_j|
\right) \;.
\label{eq:displacement}
\end{equation}
Our notation makes clear that both sides of this equation refer to the
same quantum state, but different utility operators.  In contrast, 
Eq.~(D8) is ambiguous.  The left-hand side of Eq.~(D8) refers to the 
state $\sum_j\lambda_j|x_j+k\rangle$, whereas the right-hand side refers 
to the state $\sum_j\lambda_j|x_j\rangle$; it is unclear whether the two 
sides refer to different quantum states or to a single state labeled 
according to two different utility operators.  We adopt the latter 
interpretation as being the one most consistent with Deutsch's 
discussion.  The second property of value functions, Deutsch's 
{\it zero-sum property}~(D9), becomes in our notation, 
\begin{equation}
{\cal V}\!\left( \sum_j\lambda_j|\phi_j\rangle;
\sum_j(-x_j)|\phi_j\rangle\langle\phi_j| \right)
 = - {\cal V}\!\left( \sum_j\lambda_j|\phi_j\rangle;
\sum_jx_j|\phi_j\rangle\langle\phi_j| \right)
\;.
\label{eq:zerosum}
\end{equation}
Equation~(D9) suffers from the same sort of ambiguity as Eq.~(D8): it
refers to two states, $\sum_j\lambda_j|{-x_j}\rangle$ and 
$\sum_j\lambda_j|x_j\rangle$.  In Eq.~(\ref{eq:zerosum}) we again
choose the interpretation that there is a single state, but two
different utility operators.  The zero-sum property is an axiom of 
Deutsch's nonprobabilistic decision theory, and the displacement 
property follows from the principle of additivity, another axiom of 
his analysis.

The derivation of the pivotal result deals with a state $|\psi\rangle$
that is a superposition of two utility eigenstates: 
\begin{equation}
\ket{\psi}  = \lambda_1 |\phi_1\rangle + \lambda_2 |\phi_2\rangle\;.
\end{equation}
When writing the displacement and zero-sum properties for such a state,
we can omit the other eigenstates from the utility operator, since as
Deutsch shows, the corresponding outcomes do not occur.  To shorten the
equations, we introduce the abbreviations
\begin{equation}
\hat\Pi_i = \proj{\phi_i}\;,\;\;i=1,2. 
\end{equation}

We now proceed in our notation through the rest of the argument leading 
to Eq.~(D11).  We carry along arbitrary amplitudes $\lambda_1$ and 
$\lambda_2$, because this helps to illustrate the nature of the hidden 
assumption.  The reasoning begins with the displacement property~(D8), 
specialized to the case $k=-x_1 - x_2$: 
\begin{equation}
{\cal V}\big(\ket{\psi};x_1 \hat\Pi_1 + x_2 \hat\Pi_2\big)
-x_1-x_2 
= {\cal V}\big(\ket{\psi};-x_1 \hat\Pi_2 - x_2\hat\Pi_1\big)\;.
\end{equation}
Applying the zero-sum property~(\ref{eq:zerosum}) to the game on the 
right-hand side yields
\beq
{\cal V}\big(\ket{\psi};x_2 \hat\Pi_1 + x_1 \hat\Pi_2\big) + {\cal
V}\big(\ket{\psi};x_1 \hat\Pi_1 + x_2 \hat\Pi_2\big) = x_1 + x_2 \;.
\label{eq:sum}
\eeq
Deutsch uses this result in the case $\lambda_1=\lambda_2=1/\sqrt2$,
where it becomes
\begin{equation}
{\cal V}\!\left({1\over\sqrt{2}}(\ket{\phi_1} + \ket{\phi_2});
     x_2 \hat\Pi_1 + x_1 \hat\Pi_2\right) + 
{\cal V}\!\left({1\over\sqrt{2}}(\ket{\phi_1} + \ket{\phi_2});
    x_1 \hat\Pi_1 + x_2 \hat\Pi_2\right)=
x_1+x_2\;.
\label{eq:dcase}
\end{equation}
In Deutsch's notation the values of the two games in this equation are 
denoted in the same way, so he assumes they are equal, 
\begin{equation}
{\cal V}\!\left({1\over\sqrt{2}}(\ket{\phi_1} + \ket{\phi_2});
     x_2 \hat\Pi_1 + x_1 \hat\Pi_2\right)  
= {\cal V}\!\left({1\over\sqrt{2}}(\ket{\phi_1} + \ket{\phi_2});
    x_1 \hat\Pi_1 + x_2 \hat\Pi_2\right)\;,
\label{eq:hiddenass}
\end{equation}
which leads immediately to the pivotal result~(\ref{eq:pivotal}).

Equation~(\ref{eq:hiddenass}) is the hidden assumption in Deutsch's 
argument.  To see that it is required and that it involves introducing 
the notion of probabilities, consider the following rule for measurement
outcomes: the result associated with eigenstate $|\phi_1\rangle$ 
{\it always\/} occurs.  This deterministic rule is perfectly legitimate 
at this point in the argument.  In Eq.~(\ref{eq:dcase}) it gives 
utility $x_2$ in the first game and utility $x_1$ in the second, 
thus satisfying the equation.  

Another way to get at the import of Deutsch's hidden assumption is
to make a similar assumption for the case of arbitrary expansion 
coefficients, that is to assume
\begin{equation}
{\cal V}\big(\lambda_1 \ket{\phi_1} + \lambda_2 \ket{\phi_2};
x_2 \hat\Pi_1 + x_1 \hat\Pi_2\big)
= {\cal V}\big(\lambda_1 \ket{\phi_1} + \lambda_2
\ket{\phi_2};x_1 \hat\Pi_1 + x_2 \hat\Pi_2\big)\;.
\label{eq:general}
\end{equation}
Both this assumption and the more specialized one embodied in 
Eq.~(\ref{eq:hiddenass}) are equally well (or badly) justified at this
stage of the argument.  The reason is that as yet $\lambda_1$ and 
$\lambda_2$ are just numbers attached to the possible outcomes, having
no {\it a priori\/} relation to probabilities.  Substituting 
Eq.~(\ref{eq:general}) into Eq.~(\ref{eq:sum}) gives
\begin{equation}
{\cal V}\big(\lambda_1 \ket{\phi_1} + \lambda_2 \ket{\phi_2};x_1
\hat\Pi_1 + x_2 \hat\Pi_2\big) = {1\over2}(x_1+x_2)\;, 
\label{eq:equal}
\end{equation}
which generalizes to the rule that the value of a game is the 
arithmetic mean of the utilities that have nonzero amplitude.  This
corresponds to the probability rule $p_j=(\mbox{\rm number of nonzero
amplitudes})^{-1}$.  Notice that this probability rule is contextual
in the sense of Gleason's theorem (see discussion in 
Sec.~\ref{sec:conc}).

We conclude that to derive the pivotal result~(\ref{eq:pivotal}),
one must include Eq.~(\ref{eq:hiddenass}) as an additional assumption. 
In our view, including this additional assumption is not just a 
minor addition to Deutsch's list of assumptions, but rather a major 
conceptual shift.  The assumption is akin to applying Laplace's 
Principle of Insufficient Reason to a set of indistinguishable 
alternatives, an application that requires acknowledging {\it a 
priori\/} that amplitudes are related to probabilities.  Once this 
acknowledgement is made, however, the pivotal result~(\ref{eq:pivotal}) 
is a simple consequence of classical decision theory, as can be seen 
in the following way.   As discussed in Sec.~\ref{sec:assumptions}, 
the existence of a numerical value ${\cal V}(\ket{\psi};\hat X)$ for 
each game,
together with standard nonprobabilistic axioms of decision theory,
entails that there exist probabilities $p_1$ and $p_2$ such 
that
\begin{equation}
{\cal V}\!\left({1\over\sqrt{2}}(\ket{\phi_1} + \ket{\phi_2});
     x_1 \hat\Pi_1 + x_2 \hat\Pi_2\right) 
= F^{-1}\!\big(p_1F(x_1)+p_2F(x_2)\big)\;,
\end{equation}
where $F$ is a strictly increasing function.  The hidden 
assumption~(\ref{eq:hiddenass}) then takes the form
\begin{equation}
p_1F(x_2) + p_2F(x_1) = p_1F(x_1)+p_2F(x_2) \;.
\end{equation}
If this is to be true for arbitrary $x_1$ and $x_2$ (or for any
$x_1\ne x_2$), it follows that $p_1=p_2=1/2$.  Thus in the context of
classical decision theory, the assumption (\ref{eq:hiddenass}) is
equivalent to applying the Principle of Insufficient Reason to the
case of equal amplitudes.

It is difficult to assess the validity of Deutsch's argument once one
gets past the derivation of the pivotal result.  This is mainly 
because the remainder of the argument repeatedly invokes the principle 
of substitutability.  The difficulty is that this principle---that 
the value of a game is unchanged when a subgame is replaced by another 
subgame of equal value---is never given a precise mathematical 
formulation in the quantum context.  In any case, the remainder of 
the argument can be simplified once one realizes that the Principle 
of Insufficient Reason is an essential ingredient, for then one gets 
immediately that for an equal superposition of $n$ eigenstates, the 
probability of each outcome is $1/n$.

The vagueness of the principle of substitutability has an important 
consequence.  We believe that the probability rule following
Eq.~(\ref{eq:equal}) satisfies all of Deutsch's assumptions, including
a suitably defined principle of substitutability.  If it
does, then it shows that no amount of cleverness in using Deutsch's
assumptions can ever lead uniquely to the standard quantum rule for
probabilities.  The fly in the ointment is that without a precise 
formulation of the principle of substitutability, it is not possible
to tell whether this rule satisfies it.

\section{Conclusion}
\label{sec:conc}

We have seen that if one assumes the nonprobabilistic part of 
classical decision theory, then one is effectively introducing probabilities at
the same time.  Indeed, once one realizes that quantum theory deals
with uncertain outcomes, one is forced to introduce probabilities, 
as they provide the {\it only\/} language for quantifying 
uncertainty \cite{Savage72,Caves00,deFinetti72,Cox46}.  From this
point of view, the most powerful and compelling derivation of the 
quantum probability rule is Gleason's theorem. 

\begin{quote}
{\bf Gleason's theorem:} \cite{Gleason57,Cooke81,Pitowsky98}
Assume there is a function $f$ from the one-dimensional projectors 
acting on a Hilbert space of dimension greater than 2 to the unit 
interval, with the property that for each orthonormal basis 
$\{|\psi_k\rangle\}$,
\begin{equation}
\sum_k f\big(|\psi_k\rangle\langle\psi_k|\big) = 1 \;.
\end{equation}
Then there exists a density operator $\hat\rho$ such that
\begin{equation}
f\big(|\psi\rangle\langle\psi|\big) = \langle\psi|\hat\rho|\psi\rangle \;.
\end{equation}
\end{quote}

\noindent

It is worthwhile to ponder the meaning of this theorem.
It assumes the Hilbert-space structure of observables---that
is, that each orthonormal basis corresponds to the mutually exclusive 
results of a measurement of some observable.  It sets as its task 
to derive the probabilities for the inevitably uncertain measurement
outcomes.  The only further ingredient required is that the probability 
for obtaining the result corresponding to a normalized vector 
$|\psi\rangle$ depends only on $|\psi\rangle$ itself, not on 
the other vectors in the orthonormal basis defining a particular 
measurement.  This important assumption, which might be called the
``noncontextuality'' of the probabilities, means that the probabilities 
are consistent with the Hilbert-space structure of the observables.
With these assumptions the probabilities for all measurements can 
be derived from a density operator $\hat\rho$ using the standard quantum
probability rule.  Remarkably this conclusion does not rely on 
any assumption about the continuity or differentiability of $f$;  
the only essential property of $f$ is that it be bounded. 

{\it By assuming that measurements are described by probabilities 
that are consistent with the Hilbert-space structure of the 
observables, Gleason's theorem derives in one shot the state-space
structure of quantum mechanics and the probability rule.}  It is
hard to imagine a cleaner derivation of the probability rule than 
this. 

\begin{acknowledgements}
This work was supported in part by the 
the U.S. Office of Naval Research (Grant No.~N00014-1-93-0116)
and the U.S. National Science Foundation (Grant No.~PHY-9722614). 
H.B.\ is partially supported by the Institute for Scientific 
Interchange Foundation (I.S.I.) and Elsag-Bailey, and C.A.F.\ 
acknowledges the support of the Lee A.~DuBridge Prize Postdoctoral 
Fellowship at Caltech.  C.M.C., C.A.F., and R.S.\ thank the Isaac Newton 
Institute for its hospitality, H.B.\ thanks the I.S.I. for its hospitality, 
and J.F.\ acknowledges the hospitality 
of Lawrence Berkeley National Laboratory.  
\end{acknowledgements}

\end{document}